\documentclass[preprint,secnumarabi,superscriptaddress,amssymb,nobibnotes,prd,preprintnumbers]{revtex4}

\usepackage{epsfig}
\usepackage{amsmath}
\usepackage{graphicx}
\usepackage{hyperref}

\def\p{I\!\!P}

\begin{document}

\title{How large is the diffractive contribution to inclusive dijet photoproduction in UPCs at the LHC?}

\author{Vadim Guzey}

\affiliation{National Research Center ``Kurchatov Institute'', Petersburg
 Nuclear Physics Institute (PNPI), Gatchina, 188300, Russia}
 
\author{Michael Klasen}
 
\affiliation{Institut f\"ur Theoretische Physik, Westf\"alische Wilhelms-Universit\"at M\"unster, Wilhelm-Klemm-Stra\ss{}e 9, 48149 M\"unster, Germany}



\begin{abstract}

Using next-to-leading order (NLO) perturbative QCD, we calculate the diffractive contribution to inclusive dijet photoproduction in Pb-Pb ultraperipheral collisions (UPCs) at the LHC and find that it does not exceed $5-10$\% 
in small-$x_A$ bins in the ATLAS kinematics at $\sqrt{s_{NN}}=5.02$ TeV.
Its smallness is a result of the restricted kinematics ($p_{T1} > 20$ GeV and $x_A > 0.001$), 
the large nuclear suppression of nuclear diffractive parton distribution functions predicted in the leading twist model of nuclear shadowing,
and additional suppression due to QCD factorization breaking in diffraction.
At the same time, using looser $p_T$ cuts, e.g., $p_{T1} > 10$ GeV and $p_{T i\neq 1} > 5-7$ GeV, we find that  
 $(d\sigma_{\rm diff}/dx_A)/(d\sigma_{\rm inc}/dx_A)$ can reach $10-20$\% at $x \approx 5 \times 10^{-4}$.
Also, 
applying our framework to proton-proton UPCs at $\sqrt{s_{NN}}=13$ TeV, we find that the ratio of the diffractive
and inclusive cross sections of dijet photoproduction 
$(d\sigma_{\rm diff}/dx_p)/(d\sigma_{\rm inc}/dx_p)$
is as large as $10-15$\% for $x_p \sim 5 \times 10^{-4}$.
An account of the contribution of Pomeron-Pomeron scattering, which is not included in our analysis, should make this ratio 
somewhat larger.

\end{abstract}

\preprint{MS-TP-20-46}

\maketitle

\section{Introduction}
\label{sec:1}

Electroproduction and photoproduction of jets in electron-hadron scattering at high energies at Hadron-Electron Ring Accelerator (HERA) has proven to be a useful tool to study various aspects of the dynamics and structure of hadrons in Quantum Chromodynamics (QCD), for reviews, see Refs. \cite{Klasen:2002xb,Butterworth:2005aq,Newman:2013ada}. After the closure of HERA and the advent of
the Large-Hadron Collider (LHC), there has been a growing interest in so-called ultraperipheral collisions (UPCs) at the LHC allowing
one to study photon-proton and photon-nucleus scattering at previously unattainable high energies \cite{Bertulani:2005ru,Baltz:2007kq,Klein:2019qfb}.

Our study focuses on dijet photoproduction in proton-proton and heavy-ion UPCs in the LHC kinematics, which in the latter case 
had been measured by the ATLAS collaboration \cite{ATLAS:2017kwa,Angerami:2017kot}. In particular, using the framework 
of next-to-leading order (NLO) perturbative QCD applied to inclusive \cite{Guzey:2018dlm,Guzey:2019dpp} and diffractive \cite{Guzey:2016tek} dijet photoproduction in UPCs at the LHC, we quantify the diffractive contribution to inclusive dijet photoproduction and find that it does not exceed 
$5-10$\% in small-$x_A$ bins for the kinematic distributions studied by ATLAS at $\sqrt{s_{NN}}=5.02$ TeV.
This estimate is relevant for obtaining complementary constraints on nuclear parton distribution functions (nPDFs) \cite{Guzey:2019kik}
using the LHC data on inclusive dijet photoproduction in heavy-ion UPCs in this kinematics and quantifies 
the correction factor to the inclusive data due to the experimentally excluded diffractive contribution, see also the discussion in
 \cite{Baltz:2007kq,Strikman:2005yv}.

At the same time, to enhance the diffractive signal at small $x_A^{\rm obs}$, one needs to lower $p_T$ and/or
increase the invariant
collision energy. 
We demonstrate 
 it by using the $p_{T1} > 10$ GeV and $p_{T i\neq 1} > 5-7$ GeV cut and find that  
 $(d\sigma_{\rm diff}/dx_A)/(d\sigma_{\rm inc}/dx_A)=10-20$\% at $x \approx 5 \times 10^{-4}$.
 As an illustration of the effect of increasing energy, we consider
proton-proton UPCs at $\sqrt{s_{NN}}=13$ TeV and show that the ratio of the diffractive
and inclusive cross sections of dijet photoproduction can reach 
$10-15$\% for $x_p \sim 5 \times 10^{-4}$.

Note that the ratios of cross sections of diffractive and inclusive dijet photoproduction turn out to be similar for Pb and proton targets
as a consequence of large nuclear shadowing suppressing nuclear diffractive PDFs stronger than usual nPDFs \cite{Frankfurt:2011cs} such that their ratio does not show a strong nuclear dependence.

Our NLO perturbative QCD analysis of dijet photoproduction in UPCs at the LHC extends the leading-order (LO)
analyses of this process in the frameworks of collinear factorization \cite{Strikman:2005yv,Basso:2017mue} and PYTHIA Monte Carlo \cite{Helenius:2018bai,Helenius:2019gbd} and complements predictions based on the color glass condensate 
and transverse momentum factorization \cite{Kotko:2017oxg}.
It can also be used to stimulate studies of inclusive and diffractive dijet photoproduction in proton-proton and heavy-ion UPCs 
at future HL-LHC and HE-LHC \cite{Citron:2018lsq,Klasen:2019dpu}.

\section{Cross sections of inclusive and diffractive dijet photoproduction in next-to-leading order QCD}
\label{sec:2}

Our analysis of cross sections of inclusive and diffractive dijet photoproduction is based on an analytical approach in the 
framework of collinear factorization and next-to-leading order (NLO) perturbative QCD originally developed in Refs. \cite{Klasen:1995ab,Klasen:1996it}.
In this approach and using the Weizs\"acker-Williams method of equivalent photons \cite{Bertulani:2005ru,Baltz:2007kq},
the cross section of inclusive dijet photoproduction in UPCs of ions $A$ (heavy ions, protons) $A+A \to A + {\rm 2jets} + X$ 
can be written in the following form
\begin{eqnarray}
 d\sigma(A+A &\to& A + {\rm 2jets} +X) \nonumber\\
&=&\sum_{a,b} \int dy \int dx_{\gamma} \int dx_{A}\,
f_{\gamma/A}(y) f_{a/\gamma}(x_{\gamma},m_f^2) f_{b/A}(x_A,m_f^2) d\hat{\sigma}_{ab}^{(n)} \,,
\label{eq:cs_incl}
\end{eqnarray}
where $a$ and $b$ are parton flavors with $a$ including also the photon corresponding to the direct photon contribution;
$y$, $x_{\gamma}$, and $x_A$ are longitudinal momentum fractions carried by photons, partons in the photon, and partons in the target
nucleus (proton), respectively;
$f_{\gamma/A}(y)$ is the photon flux calculated in the equivalent photon approximation;
$f_{a/\gamma}(x_{\gamma},m_f^2)$ and 
$f_{b/A}(x_A,m_f^2)$ are parton distribution functions (PDFs) of the photon in the resolved photon case and hadron $A$, respectively; 
$d\hat{\sigma}_{ab}^{(n)}$ is the cross section for the
production of an $n$-parton final state from two initial partons $a$ and $b$. 
Following our findings in \cite{Guzey:2018dlm}, both photon and proton PDFs are taken at the equal
factorization scale, which in our analysis is identified with twice the average dijet transverse momentum $m_f=2\bar{p}_T=p_{T1}+p_{T2}$.
To quantify the sensitivity of our results to the choice of $m_f$, we follow the standard prescription and vary it in the interval
$(m_f/2, 2 m_f)$.

Using the experimentally detected dijet final state, one can determine hadron-level estimates for the parton momentum fractions $z_{\gamma}^{\rm obs}$ and $x_A^{\rm obs}$ in the photon and nuclear target, respectively~\cite{ATLAS:2017kwa}
\begin{equation}
z_{\gamma}^{\rm obs}=\frac{m_{\rm jets}}{\sqrt{s_{NN}}}\, e^{y_{\rm jets}} \,, \qquad x_{A}^{\rm obs}=\frac{m_{\rm jets}}{\sqrt{s_{NN}}}\, e^{-y_{\rm jets}} \,,
\label{eq:obs}
\end{equation}
where $y_{\rm jets}$ and $m_{\rm jets}$ are the dijet rapidity and mass, respectively, which can be calculated 
using the measured jet energies $E_i$, the three-momenta $\vec{p}_i$ and their longitudinal components $p_{i,z}$,
\begin{equation}
y_{\rm jets}=\frac{1}{2} \ln \left(\frac{\sum_i E_i+p_{i,z}}{\sum_i E_i-p_{i,z}}\right) \,, \qquad
m_{\rm jets}=\left[\left(\sum_i E_i\right)^2-\left(\sum_i \vec{p}_i\right)^2\right]^{1/2} \,.
\label{eq:obs2}
\end{equation}
At leading-order (LO), i.e., in the $2 \to 2$ kinematics, $x_A=x_{A}^{\rm obs}$ and $x_{\gamma} y=z_{\gamma}^{\rm obs}$ in
Eq.~(\ref{eq:cs_incl}). Moreover, for the direct-photon contribution, $x_{\gamma}=1$.
At NLO due to QCD radiative corrections, the momentum fractions in Eq.~(\ref{eq:cs_incl}) 
are generally somewhat larger than their hadron-level estimates of Eq.~(\ref{eq:obs}).

The dijet rapidity in Eqs.~(\ref{eq:obs}) and (\ref{eq:obs2}) is defined with respect to the direction of the photon-emitting ion.
The latter is unambiguously anti-correlated with the direction of the hadronic final state $X$. In the ATLAS measurement~\cite{ATLAS:2017kwa}, it is 
determined by selecting events satisfying the 0nXn condition in the zero-degree calorimeter (ZDC) corresponding to zero neutrons in one direction and one of more neutrons in the opposite direction. This fills the rapidity gap on the nuclear target side and removes
nuclear diffraction from the selected 0nXn events. As we will discuss below, it has implications for the analysis of this data in terms of nuclear parton distributions.

Requiring that the target nucleus remains intact, one can consider diffractive dijet photoproduction in $AA$ UPCs
 $A+A \to A + {\rm 2jets} + X+A$. The cross section of this process is given by a sum of two terms, which reflects 
 the possibility for each ion to serve as a source of photons and as a target (the interference between them is
 negligibly small for the considered observables)
\begin{eqnarray}
d\sigma(A&+&A \to A + {\rm 2jets} +X+A) \nonumber\\
&=&d\sigma(A+A \to A + {\rm 2jets} +X+A)^{(+)}+d\sigma(A+A \to A + {\rm 2jets} +X+A)^{(-)} \,.
\label{eq:sym_diff}
\end{eqnarray}
The two contributions in Eq.~(\ref{eq:sym_diff}) are connected by inversion of the signs of the jet rapidities, which
corresponds to inversion of the direction of the photon-emitting nucleus.

By analogy with the inclusive case, the $d\sigma(A+A \to A + {\rm 2jets} +X+A)^{(+)}$
cross section can be calculated through
\begin{eqnarray}
 d\sigma(A+A & \to & A + {\rm 2jets} +X+A)^{(+)}= \sum_{a,b} \int dy \int dx_{\gamma} \int dt \int dx_{\p} \int dz_{\p} \nonumber\\
 & \times & f_{\gamma/A}(y) f_{a/\gamma }(x_\gamma,m_f^2) f^{D(4)}_{b/A}(x_{\p},t,z_{\p},m_f^2) d\hat{\sigma}_{ab}^{(n)} \,,
  \label{eq:cs_diff}
  \end{eqnarray}
where $x_{\p}$ and $z_{\p}$ refer to the momentum fraction of the target nucleus (proton) carried by the diffractive exchange (Pomeron and Reggeon)
and the parton momentum fraction in the Pomeron, respectively; $t$ is the invariant momentum transfer squared;
$f^{D(4)}_{b/A}$ denotes the diffractive PDF of the target. 
Note that since QCD factorization for diffractive dijet photoproduction is broken~\cite{ZEUS:2007uvk,H1:2007jtx,H1:2010xdi,H1:2015okx},
 Eq.~(\ref{eq:cs_diff}) also requires introduction of a model-dependent suppression factor (rapidity gap survival probability)~\cite{Klasen:2008ah,Klasen:2010vk,Guzey:2016awf}
 (see the details below).
 
In our numerical analysis of Eqs.~(\ref{eq:cs_incl}) and (\ref{eq:cs_diff}), we used the following input.
The flux of equivalent photons is given by convolution over the impact parameter $\vec{b}$ 
of the flux of quasireal photons emitted by an ultrarelativistic charged ion 
$N_{\gamma/A}(y,{\vec b})$~\cite{Budnev:1975poe,Vidovic:1992ik} with the probability not to have inelastic strong ion-ion interactions 
$\Gamma_{AA}({\vec b})=\exp[-\sigma_{NN} \int d^2 {\vec b^{\prime}} T_A({\vec b^{\prime}}) T_A({\vec b} -{\vec b^{\prime}})]$, 
\begin{equation}
f_{\gamma/A}(y)=\int d^2 {\vec b}\, N_{\gamma/A}(y,\vec{b}) \Gamma_{AA}({\vec b}) \,,
\label{eq:flux_dop}
\end{equation}
where $\sigma_{NN}$ is the total nucleon-nucleon cross section; $T_A({\vec b})=\int dz \rho_A({\vec b},z)$ is the nuclear optical density with $\rho_A({\vec b},z)$ being the nuclear density.
However, for practical applications, one can show~\cite{Guzey:2016tek,Nystrand:2004vn} that this exact expression can be very well approximated by 
the flux of equivalent photons produced by
a relativistic point-like charge $Z$
\begin{equation}
f_{\gamma/A}(y)=\frac{\alpha_{\rm e.m.} Z^2}{\pi} \frac{1}{y} 
\left[2 \xi K_0(\xi) K_1(\xi)-\xi^2 \left(K_1^2(\xi)-K_0^2(\xi)\right) \right] \,,
\label{eq:flux}
\end{equation}
where $\alpha_{\rm e.m.}$ is the fine-structure constant; $K_{0,1}$ are modified Bessel functions of the second kind; 
$\xi=y m_p b_{\rm min}$ with $m_p$ the proton mass and $b_{\rm min}=14.2$ fm~\cite{Nystrand:2004vn} the minimal impact parameter between the colliding ions in Pb-Pb UPCs.

In the proton case, we use the Drees and Zeppenfeld (DZ) result for the photon flux~\cite{Drees:1988pp}
\begin{equation}
f_{\gamma/p}(y)=\frac{\alpha_{\rm e.m.}}{2 \pi} \frac{1+(1-y)^2}{y} \left[\ln A -\frac{11}{6}+\frac{3}{A}-\frac{3}{2A^2}
+\frac{1}{3A^3}
\right] \,, 
\label{eq:flux_p}
\end{equation}
where $A=1+(0.71\ {\rm GeV}^2)/Q^2_{\rm min}$ and $Q^2_{\rm min}=(y m_p)^2/(1-y)$ is the minimal kinematically allowed photon virtuality.

The resolved photon contribution involves the $f_{a/\gamma}$ photon PDFs, for which we use the GRV NLO parametrization
transformed from the DIS$_{\gamma}$ to $\overline{\rm MS}$ scheme \cite{Gluck:1991jc}.

We tested two recent NLO parametrizations of nuclear PDFs $f_{b/A}$, namely nCTEQ15 \cite{Kovarik:2015cma} and EPPS16 \cite{Eskola:2016oht};
the $f_{b/p}$ proton PDFs  are also taken from the nCTEQ15 fit.

In the case of diffractive dijet photoproduction on the proton, our calculations involve the diffractive PDFs of the proton $f^{D(4)}_{b/p}$, which have been determined from QCD analyses of inclusive diffraction in $ep$ DIS at HERA \cite{Aktas:2006hy,Chekanov:2009aa,Goharipour:2018yov,Khanpour:2019pzq} with an addition in some cases of the
data on diffractive dijet production \cite{Chekanov:2009aa,Aktas:2007bv}.
Based on these studies, we will use the 2006 H1 Fit B \cite{Aktas:2006hy} as our base parametrization of
$f^{D(4)}_{b/p}$
 since 
it agrees well with the inclusive and dijet H1 and ZEUS HERA data~\cite{Aktas:2006hy,Chekanov:2009aa,Aktas:2007bv}
 as well as with the recent independent 
analyses~\cite{Goharipour:2018yov,Khanpour:2019pzq}.

Nuclear diffractive PDFs $f^{D(4)}_{b/A}$ are novel distributions that have never been measured and, hence, one must rely on their modeling.
We used the leading twist model of nuclear shadowing proposing a microscopic mechanism of nuclear suppression (shadowing)
of nuclear diffractive PDFs \cite{Frankfurt:2011cs}. In this approach, in a wide range of the $x_{\p}$ and $z_{\p}$ momentum 
fractions and the resolution scale $m_f$, nuclear effects weakly depend on these variables and the parton flavor $b$ and, to a good accuracy, one has (see Ref. \cite{Guzey:2016tek})
\begin{equation}
f^{D(4)}_{b/A}(x_{\p},t,z_{\p},m_f^2)= R_b\, A^2 F_A^2(t) f^{D(4)}_{b/p}(x_{\p},t_{\rm min},z_{\p},m_f^2) \,,
\label{eq:dpdfs_A}
\end{equation}
where $R_b = 0.1-0.2$ is the nuclear suppression factor characterizing the predicted strong nuclear shadowing effect;
$F_A(t)$ is the nuclear form factor; $t_{\rm min}=-(x_{\/p}m_p)^2/(1-x_{\p})$ is the minimal momentum transfer squared.
The spread in the value of $R_b$ reflects the significant uncertainty 
in predictions of the leading twist model of nuclear shadowing
for nuclear diffractive PDFs.

We mentioned above that analyses of diffractive dijet photoproduction in electron-proton scattering at HERA indicated that 
QCD factorization in this process is broken, i.e., NLO pQCD calculations overestimate the measured cross sections by almost a factor
of two~\cite{ZEUS:2007uvk,H1:2007jtx,H1:2010xdi,H1:2015okx,Klasen:2008ah,Klasen:2010vk}. The pattern of this factorization breaking 
is not firmly established: NLO pQCD provides a consistent description of the data after introducing either the global suppression factor of $R(\rm glob.)=0.5$ or the suppression factor of $R(\rm res.)=0.34$~\cite{Kaidalov:2003xf,Kaidalov:2009fp} only for the resolved photon contribution.  
In addition, the suppression factor may depend on the parton flavor and the $x_{\gamma}$ momentum fraction~\cite{Guzey:2016awf}.

In the case of diffractive production off nuclei, the magnitude of factorization breaking is expected to be larger than that in the 
proton case because soft inelastic interactions leading to population of the rapidity gap are enhanced in the nuclear case. 
For instance, using the two-state eikonal model~\cite{Kaidalov:2003xf,Kaidalov:2009fp}, one can estimate that 
$R(\rm res.) \approx 0.04$ for the lead target~\cite{Guzey:2016tek}.
The effects of factorization breaking for the nuclear and proton targets will be included in our numerical results presented in the following section.

\section{Predictions for dijet photoproduction in heavy-ion and proton-proton UPCs at the LHC}
\label{sec:3}

Our numerical calculations of cross sections of inclusive and diffractive dijet photoproduction using Eqs. (\ref{eq:cs_incl}),
 (\ref{eq:sym_diff})
and (\ref{eq:cs_diff}) with the input discussed in the previous sections are performed using the NLO parton-level Monte Carlo 
framework developed in Refs. \cite{Klasen:1995ab,Klasen:1996it,Klasen:2010vk}.
It implements the kinematic conditions and cuts used in the ATLAS analysis \cite{ATLAS:2017kwa,Angerami:2017kot}, 
namely, the anti-$k_T$ algorithm with the jet radius $R = 0.4$;
the invariant collision energy per nucleon is $\sqrt{s_{NN}}=5.02$ TeV;
the leading jet with $p_{T1} > 20$ GeV and all other jets with $p_{T i\neq 1} > 15$ GeV corresponding to $35 < H_{T} < 400$ GeV, where $H_{T}=\sum_i p_{Ti}$; the rapidities of all jets are within the $|\eta_i| < 4.4$ interval; 
the combined mass of all reconstructed jets is $35 < m_{\rm jets} < 400$ GeV;
the parton momentum fraction on the photon side $z_{\gamma}=yx_{\gamma}$ is within the $10^{-4} < z_{\gamma} < 0.05$ interval;
the parton momentum fraction on the nucleus side $x_A$ is restricted by the $5 \times 10^{-4} < x_A < 1$ condition.

\begin{figure}[t]
\centering
 \epsfig{file=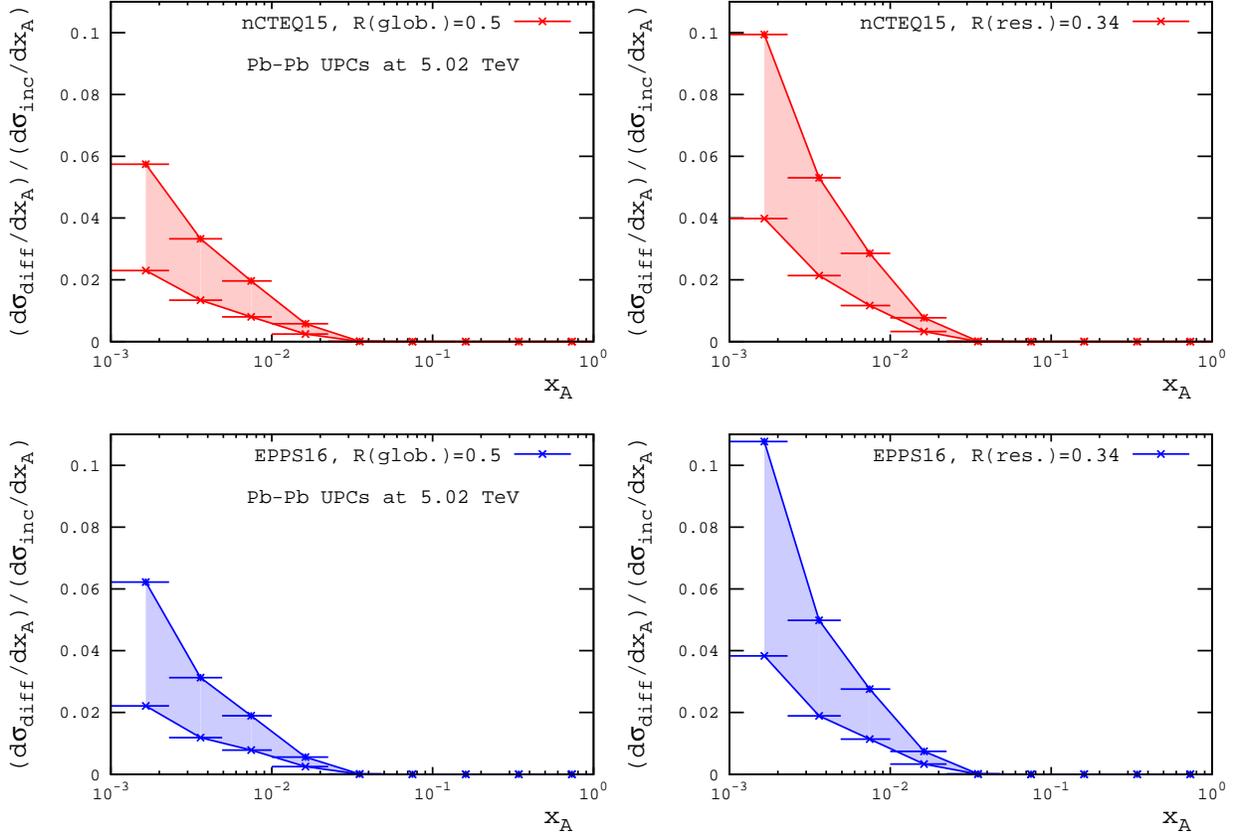,width=1.0\textwidth}
\caption{The ratio of the cross sections of diffractive and
inclusive dijet photoproduction in Pb-Pb UPCs at $\sqrt{s_{NN}}=5.02$ TeV in the ATLAS kinematics as a function of 
$x_A$. 
The upper and lower panels correspond to nCTEQ15 and EPPS16 nPDFs, respectively; the error bands quantify 
the uncertainty 
of our theoretical predictions including the uncertainties in the magnitude of nuclear shadowing in nuclear diffractive PDFs,
the uncertainties of nCTEQ15 and EPPS16 nPDFs, and the uncertainty in the choice of the scale $m_f$.
The left and right panels assume two different scenarios of QCD factorization breaking based either on the global suppression factor of $R(\rm glob.)=0.5$ or on $R(\rm res.)=0.34$ for the resolved photon contribution.
}
\label{fig:1}
\end{figure}

Among several possible kinematic distributions of dijet photoproduction, see Ref.~\cite{Guzey:2018dlm}, 
the $x_A$ dependence reveals best the
role of the diffractive contribution.  Figure \ref{fig:1} shows our predictions for the 
$(d\sigma_{\rm diff}/dx_A)/(d\sigma_{\rm inc}/dx_A)$
ratio of the cross sections of diffractive and
inclusive dijet photoproduction in Pb-Pb UPCs at $\sqrt{s_{NN}}=5.02$ TeV in the kinematic setup presented above as a function of 
$x_A$. The upper and lower panels correspond to the calculation using the nCTEQ15 and EPPS16 nPDFs, respectively.
 The error bands reflect the uncertainty of our theoretical predictions 
 and include (in the order of importance) the uncertainty in the magnitude of nuclear shadowing in nuclear diffractive PDFs 
 quantified by the factor of $R_b = 0.1-0.2$, the uncertainties of nCTEQ15 and EPPS16 nPDFs calculated using the corresponding error PDFs, and the uncertainty associated with the variation of the scale $m_f$ in the interval $(m_f/2, 2 m_f)$. Note that the latter contribution largely cancels in the ratio of the diffractive and inclusive dijet cross sections.

In the presented results, we test two scenarios of QCD factorization breaking in diffraction (see the discussion above):
the left column corresponds to the suppression of the NLO pQCD results by the global suppression factor of $R(\rm glob.)=0.5$, while
the right column corresponds to the suppression of the resolved photon contribution by the factor of  $R(\rm res.)=0.34$.
We checked that in the latter scenario the use of the much stronger suppression factor $R(\rm res.)=0.04$,
which is characteristic of heavy nuclear targets, leads to very similar predictions. Thus, the magnitude of factorization breaking does not depend on the type of the target (proton or heavy nucleus). This is natural since the small $x_A$ region, where the diffractive contribution is sizable, corresponds to large $x_{\gamma}$, which is dominated by the unsuppressed direct photon contribution.

Note that as usual in the case of coherent (i.e.~without nuclear break-up) photoproduction in Pb-Pb UPCs, the diffractive 
cross section receives contributions from both left-moving and right-moving sources of quasireal photons, see Eq.~(\ref{eq:sym_diff}).
This leads to a symmetric distribution in $y_{\rm jets}$ and increases the resulting cross section. In particular, it doubles 
the cross section at $y_{\rm jets}=0$ and also after integration over $y_{\rm jets}$. However, this enhancement of the diffractive contribution is either fully compensated by the reduction due to $R(\rm glob.)=0.5$ or compensated at large $x_A$ by the factor
$R(\rm res.)=0.34$.

One can see from the figure that the diffractive contribution is only sizable in first few bins in $x_A$, where it does not 
exceed $5-10$\%. Its smallness is a result of the restricted kinematics of rather large $p_{T1} > 20$ GeV and not sufficiently small
$x_A > 0.001$, the large nuclear suppression factor of $R_b = 0.1-0.2$ of nuclear diffractive PDFs  
characteristic for the leading twist model of nuclear shadowing, 
and the additional suppression because of the factorization breaking. 
This quantifies the magnitude of the correction relevant for an analysis of nPDFs at small $x_A$ using the ATLAS data on inclusive dijet photoproproduction in Pb-Pb UPCs at the LHC.
Indeed, since the inclusive cross section by definition also includes the diffractive contribution,
one has to correct the ATLAS data for it because diffraction has been experimentally excluded by selecting the 0nXn topology.
Without this correction, one would somewhat underestimate the effect of nuclear shadowing in nPDFs using the ATLAS data.

Note that the small-$x_A$ region, which is at the focus of our interest, corresponds to large values of $x_{\gamma}$ dominated by the direct photon contribution. Thus, our predictions for the cross section ratios presented in Fig. \ref{fig:1} depend 
very weakly 
on the choice of the photon PDFs and practically coincide with the ratio of the direct photon contribution to the diffractive and inclusive dijet cross sections.

To enhance the diffractive contribution, one primarily needs to lower the values of probed $x_A$, which can be readily achieved
by loosening the cut on $p_T$ and/or increasing the invariant collision energy $\sqrt{s_{NN}}$. In particular, we found that 
by using  
the $p_{T1} > 10$ GeV and $p_{T i\neq 1} > 5-7$ GeV cut, one can probe the dijet photoproduction cross section down to 
$\sim 5 \times 10^{-4}$, where the studied cross section ratio reaches $(d\sigma_{\rm diff}/dx_A)/(d\sigma_{\rm inc}/dx_A)=10-20$\%.

\begin{figure}[t]
\centering
 \epsfig{file=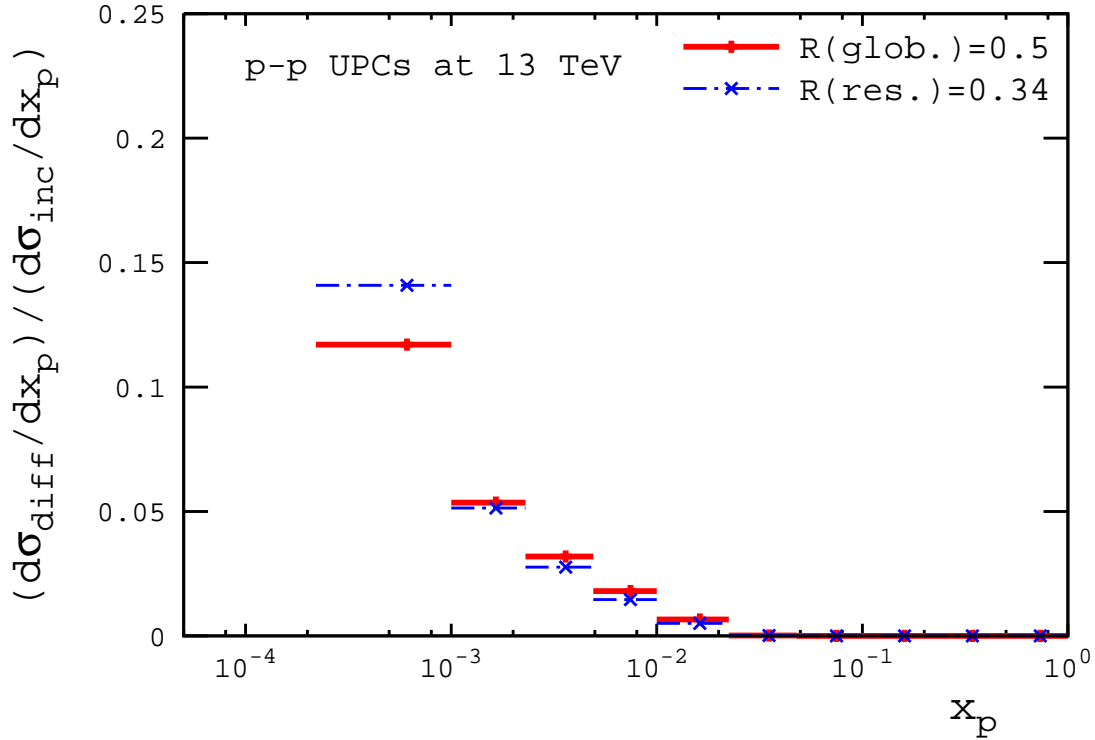,width=0.9\textwidth}
\caption{The ratio of the cross sections of diffractive and
inclusive dijet photoproduction in proton-proton UPCs at $\sqrt{s_{NN}}=13$ TeV as a function of $x_p$.
The red solid and blue dot-dashed lines correspond to $R(\rm glob.)=0.5$ and $R(\rm res.)=0.34$ assumptions on QCD factorization breaking, respectively.
}
\label{fig:2}
\end{figure}

Alternatively, one can increase the collision energy.
In the case of proton-proton UPCs, this is illustrated in Fig. \ref{fig:2}, which shows
the ratio of the cross sections of diffractive and
inclusive dijet photoproduction in proton-proton UPCs at $\sqrt{s_{NN}}=13$ TeV as a function of $x_p$.
In our calculations, we used the CTEQ15 PDFs of the free proton \cite{Kovarik:2015cma}, the 2006 H1 Fit B diffractive PDFs 
of the proton~\cite{Aktas:2006hy}, and the two scenarios of factorization breaking discussed above.
 One can see from the figure 
that it allows one to probe $x_p$ down to approximately $x_p=5 \times 10^{-4}$, where the diffractive contribution reaches 
$10-15$\%
of the inclusive dijet cross section.

Note also that the ratios of cross sections of diffractive and inclusive dijet photoproduction are similar in similar kinematics for Pb and proton targets. This is a consequence of large nuclear effects in the leading twist model of nuclear shadowing, which 
predicts that nuclear diffractive PDFs are suppressed at small $x_A$ stronger than usual (inclusive) nPDFs such that their ratio
does not show a strong nuclear dependence, see Figs.~69 and 70 of \cite{Frankfurt:2011cs}.

Our analysis relies
on the dominance of the electromagnetic (photon-Pomeron fusion) mechanism of dijet photoproduction over the competing 
Pomeron-Pomeron and photon-photon fusion mechanisms.
In the case of Pb-Pb UPCs,  this dominance 
has been confirmed by 
the analysis of Ref.~\cite{Basso:2017mue} using Forward Physics Monte Carlo (FPMC)~\cite{Boonekamp:2011ky} and is based primarily on high fluxes of equivalent photons emitted by heavy ions.
In the case of proton-proton UPCs, photon-Pomeron fusion dominates only at forward rapidities, while Pomeron-Pomeron scattering
gives the main contribution to the UPC cross section at central rapidities. 
Thus, we expect that an account of the latter contribution will somewhat increase the predicted ratio
$(d\sigma_{\rm diff}/dx_{p})/(d\sigma_{\rm inc}/dx_{p})$.
In data analysis, combining the predicted different patterns of the $y_{\rm jets}$
dependence of the competing contributions~\cite{Basso:2017mue} with the presence of additional diffractively-produced hadrons, one can in principle separate the photon-Pomeron and Pomeron-Pomeron contributions.

\section{Conclusion}
\label{sec:4}

Within NLO perturbative QCD, we calculated the diffractive contribution to inclusive dijet photoproduction in Pb-Pb UPCs
at the LHC and found that it does not exceed $5-10$\% in small-$x_A$ bins in the ATLAS kinematics at $\sqrt{s_{NN}}=5.02$ TeV.
Its smallness is a result of the restricted kinematics ($p_{T1} > 20$ GeV and $x_A > 0.001$), 
the large nuclear suppression of nuclear diffractive PDFs predicted in the leading twist model of nuclear shadowing,
and the addition suppression because of QCD factorization breaking.
 At the same time,  
 the diffractive contribution can be enhanced by lowering the $p_{T}$ cut or increasing the collision energy. For example,
 using $p_{T1} > 10$ GeV and $p_{T i\neq 1} > 5-7$ GeV cut, we find that  
 $(d\sigma_{\rm diff}/dx_A)/(d\sigma_{\rm inc}/dx_A)=10-20$\% at $x \approx 5 \times 10^{-4}$.
 Applying our framework to proton-proton UPCs at $\sqrt{s_{NN}}=13$ TeV, we found
 that the ratio of the diffractive
and inclusive cross sections of dijet photoproduction can reach $10-15$\% for $x_p \sim 5 \times 10^{-4}$.
In practice, the $(d\sigma_{\rm diff}/dx_{p})/(d\sigma_{\rm inc}/dx_{p})$ ratio should be somewhat larger 
due to the contribution of
Pomeron-Pomeron scattering, which is not included in our analysis.

\section*{Acknowledgements}
\noindent

The authors thank M.~Strikman for stimulating this study and E.~Aschenauer for the discussion of the diffractive 
contribution to the total lepton-proton DIS cross section at HERA.
M.K. would like to thank the Petersburg Nuclear Physics Institute (PNPI), Gatchina, for the kind hospitality.
V.G.'s research is supported in part by RFBR, Research Project No.~17-52-12070. 
The authors gratefully acknowledge financial support of DFG through Grant No.~KL 1266/9-1
within the framework of the joint German-Russian project ''New constraints on nuclear parton distribution functions
at small $x$ from dijet production in $\gamma A$ collisions at
the LHC''.

\end{document}